\documentstyle[12pt]{article}
\begin{document}
\thispagestyle{empty}
\begin{center}
\LARGE \tt \bf {Metric Perturbations in Einstein-Cartan Cosmology}
\end{center}

\vspace{5cm}

\begin{center}
{\large By L.C. Garcia de Andrade\footnote{Departamento de F\'{\i}sica Te\'{o}rica - Instituto de F\'{\i}sica - UERJ 

Rua S\~{a}o Fco. Xavier 524, Rio de Janeiro, RJ

Maracan\~{a}, CEP:20550-003.}}
\end{center}

\begin{abstract}
Metric perturbations the stability of solution of Einstein-Cartan cosmology (ECC) are given.
The first addresses the stability of solutions of Einstein-Cartan (EC) cosmological model against 
Einstein static universe background. In this solution we show that the metric is stable against first-order perturbations and correspond to acoustic oscillations. The second example deals with the stability of de Sitter metric also against first-order perturbations. Torsion and shear are also computed in these cases. The resultant perturbed anisotropic spacetime with torsion is only de Sitter along one direction or is unperturbed along one direction and perturbed against the other two. Cartan torsion contributes to the frequency of oscillations in the model. Therefore gravitational waves could be triggered by the spin-torsion scalar density . 
\end{abstract}

\newpage
\section{Introduction}

Scalar and metric (gravitational waves) perturbations have been extensively investigated recently \cite{1,2} in the framework of alternative theories of gravity such as Brans-Dicke and Einstein-Cartan \cite{3,4} theories. Metric perturbations have also proved very useful in the investigation of the stability of solutions of higher-order string gravity against de Sitter spacetime \cite{2}. In this case de Sitter metric was proved to be unstable but as pointed out by Maroto and Shapiro by a proper choice of parameters stable solutions could become unstable and vice-versa. Also recently stability of G\"{o}del spacetime have been investigated in the context of Einstein-Cartan cosmology \cite{5} by considering different stabilities along different axis. In this paper we consider the investigation of de Sitter metric stability and Einstein solution stability of EC field equations of gravity in its simplest form namely without dilatons and magnetic fields. Recently several types of anisotropic solutions have been considered by Lu et al. \cite{6}.
They basically considered Bianchi type $I-VII_{0}$ and Bianchi type V spacetime solutions of EC field equations with magnetic moments. In this paper we also show that de Sitter and Einstein solutions of EC gravity can be obtained from strong constraints in the spin, mass and pressure densities. This is the first step to compute metric perturbations against the original background anisotropic EC solutions. A anisotropic metric is obtained which is de Sitter only along one direction while the others are not de Sitter. It is also shown that the metric perturbation obeys a forced oscilator equation which lead to simple interpretation that the gravitational waves maybe triggered by perturbations. It is also suggested that primordial torsion could be detected as imprints in gravitational waves or in CMBR in future. This is in agreement with the idea that torsion plays an important role in the Early Universe. Recently we noticed that Cappozziello et al have worked in density perturbations within the context of torsion theories of gravity \cite{7}. D.Palle \cite{8} has also computed density perturbations and show that the Einstein-Cartan cosmology fits well within COBE data. Later on we extend Palle's work to deal with inflaton fields and torsion \cite{9,10}. Cappoziello et al approach is very interesting since it is gauge invariant. The investigation of metric perturbations is important in other direction which is in the investigation of the CMBR. Since as we shall demonstrate metric perturbations depend upon spin-torsion in second order (though can be first order in the presence of magnetic fields) the temperature anisotropy depends on torsion indirectly \cite{11}. This is in agreement with recently suggestions that torsion could be detected as a relic imprint in CMBR.  Einstein-Cartan cosmology and metric perturbations may together also be useful in the investigation of neutrinos and torsion waves in the Universe. This paper follows a series of previous papers on the study of physical aspects of ECC such as the thermodynamics in ECC \cite{9} and in the investigation of cosmological rotational perturbations (vector perturbations) in ECC \cite{11}. The paper is organized as follows. In section 2 we provide a brief review of EC gravity and spinning fluids. Section 3 we deal with the stability of de Sitter metric in ECC and the Einstein static universe solution of EC field equations with the respective perturbation. In both examples we show that the Einstein and de Sitter metric are stable in this framework.
\section{Weyssenhoff fluids in Einstein-Cartan cosmology}
Since we are not interested here in magnetized fluids we consider the energy-momentum tensor in the form
\begin{equation}
T^{i}_{j}= -p{\delta}^{i}_{j}+ u^{i}u_{j}({\rho}+p)+k S^{i}_{j}+2u^{p}{\nabla}_{k}S^{k}_{lj} 
\label{1}
\end{equation}
where $(i,j=0,1,2,3)$ and ${\nabla}_{i}$ denotes the Riemann-Cartan covariant derivative and ${\rho}$ and p are respectively the matter density and pressure of the spinning fluid and $k$ is the coupling constant between matter and spin. Here the tensor $S^{i}_{jk}$ is the spin density tensor which obeys the Frenkel condition $S^{i}_{jk}= u^{i}S_{jk}$ where $u^{i}$ is the four-velocity. Here $S_{ij}$ is the spin angular momentum tensor. 
\section{De Sitter and Einstein universes perturbations in EC cosmology}
The line element describing the anisotropic spacetime is given by
\begin{equation}
ds^{2}= dt^{2}- A^{2}(t)dx^{2} -B^{2}(t)dy^{2}- C^{2}(t)dz^{2}
\label{2}
\end{equation}
which describes a Bianchi type I Universe. We shall addopt here the constraint that $S^{12}=S^{13}=S^{23}=\frac{S(t)\sqrt{3}}{3}$. Before we consider the metric perturbations around de Sitter background we need to demonstrate that de Sitter metric itself satisfies the Einstein-Cartan equation. This can easily be proved by substitution of de Sitter metric into EC equations which reduce EC field equations to just two equations
\begin{equation}
-3H^{2}_{0}=(p-S^{2})
\label{3}
\end{equation}
\begin{equation}
3H^{2}_{0}=({\rho}-S^{2})
\label{4}
\end{equation}
which together imply the following constraint
\begin{equation}
{\rho}+p = 2S^{2}
\label{5}
\end{equation}
Note that for example the vacuum inflationary constraint ${\rho}+p=0$ would imply the vanishing of the scalar spin density $S^{2}$ is forced to vanish. However one should note that this constraint does not apply to the perturbed solution around the background de Sitter metric. To investigate the metric perturbations let us consider the following line element
\begin{equation}
ds^{2}= dt^{2}- a^{2}(t)({\delta}_{{\alpha}{\beta}} - h_{{\alpha}{\beta}})dx^{\alpha}dx^{\beta}
\label{6}
\end{equation}
where $h_{{\alpha}{\beta}}$ is the metric perturbation and ${\delta}_{{\alpha}{\beta}}$ is the three-dimensional kronecker delta where $({\alpha},{\beta}=1,2,3)$. Here $a^{2}$ represents the background metric. Let us now consider the de Sitter background metric in the form $a^{2}=e^{2H_{0}t}$ where $H_{0}= \frac{\dot{a}}{a}$ is the de Sitter factor. To cast the new perturbed matric into a anisotropic form we chooseTo further  simplify computations we shall consider that the metric (\ref{2}) shall be written the form
\begin{equation}
A^{2}= a^{2}(t)(1 - h_{11})dx^{i}dx^{j}
\label{7}
\end{equation}
\begin{equation}
B^{2}= a^{2}(t)(1 - h_{22})dx^{i}dx^{j}
\label{8}
\end{equation}
\begin{equation}
C^{2}= a^{2}(t)(1 - h_{33})dx^{i}dx^{j}
\label{9}
\end{equation}
To further simplify computations we shall consider the following choice
\begin{equation}
h_{33}=0
\label{10}
\end{equation}
and 
\begin{equation}
h=h_{11}=h_{22}
\label{11}
\end{equation}
Note that this perturbations would be isotropic only in two directions and greatly simplify our computations since in this case the $A(t)$ and $B(t)$ metric coefficients coincide and besides $C^{2}=a^{2}$ is the de Sitter coefficient of the background metric. These choices allows us to write the Einstein-Cartan equations in the form 
\begin{equation}
\frac{2\ddot{B}}{B}+{(\frac{\dot{B}}{B})}^{2}= S^{2}-p
\label{12}
\end{equation}
\begin{equation}
-\frac{\ddot{B}}{B}-\frac{\ddot{a}}{a}\frac{\dot{B}}{B}-\frac{\ddot{a}}{a}\frac{\dot{B}}{B}=p-S^{2}
\label{13}
\end{equation}
\begin{equation}
3{(\frac{\dot{B}}{B})}^{2}= {\rho}- S^{2}
\label{14}
\end{equation}
Addition of expressions (\ref{13}) and (\ref{14}) yields
\begin{equation}
\frac{\ddot{B}}{B}-\frac{\ddot{a}}{a}\frac{\dot{B}}{B}-\frac{\ddot{a}}{a}+(\frac{\dot{B}}{B})^{2}=0
\label{15}
\end{equation}
Substitution of (\ref{12}) into (\ref{15}) yields
\begin{equation}
\ddot{B}-H_{0}\dot{B}+ {{\omega}_{0}}^{2}B=0
\label{16}
\end{equation}
where ${{\omega}_{0}}^{2}= \frac{1}{3}[{\rho}-S^{2}-H^{2}_{0}]$. Now let us substitute the values of ${\rho}= \frac{M}{B^{2}a}$ and $S^{2}=\frac{L^{2}}{B^{4}a^{2}}$ obtained from the conservation equations for spin and matter density. Substitution of the B factor in terms of the metric perturbation h yields finally the metric perturbation equation
\begin{equation}
\ddot{h}-2H_{0}\dot{h}+{\omega}^{2}h=\frac{8}{3}L^{2}[1-\frac{3}{4}H_{0}t]
\label{17}
\end {equation}
where ${\omega}^{2}=\frac{3}{2}L^{2}$ where $L^{2}$ represents the spin density strenght of the model. Solution of this equation yields
\begin{equation}
h(t)=-\frac{2}{3}\frac{(-4{\omega}^{2}+6H_{0}^{2}+3H_{0}{\omega}^{2}t)}{{\omega}^{4}}+c_{\pm}e^{(H_{0}{\pm}\sqrt{H^{2}_{0}-{\omega}^{2}}t}
\label{18}
\end{equation}
Note from this expression that the first term is an unstable mode while the last two terms maybe stable or unstable depending on the relation between $H^{2}$ and ${\omega}^{2}$. In the case of Einstein static universe or even in the Minkowskian case \cite{11} one obtains the simple forced oscillator relation for the metric perturbation
\begin{equation}
\ddot{h}+{\omega}^{2}h=\frac{8}{3}L^{2}
\label{19}
\end{equation}
which yields the following solution
\begin{equation}
h(t)=\frac{8}{3{\omega}^{2}}+c_{\pm}e^{({\pm}i{\omega}t)}
\label{20}
\end{equation}
which reveals an acoustic oscillatory behaviour for the metric perturbation. This aspect is also present in the de Sitter case since h there also possess complex functions modes. Since as it is well known from the Sachs-Wolfe formula the CMBR is proportional to the metric perturbation one must conclude that CMBR background anisotropy must have a second order contribution due to torsion. The shear can be easily seem to be given by ${\sigma}^{2}=\frac{Q^{2}e^{-2H_{0}t}}{B^{2}}$ which shows that shear decays quickly as the universe suffers rapid expansion. Gravitational waves around de Sitter backgrounds in EC cosmology may also be found from the study presented here.
\section*{Acknowledgements}
 Thanks are due to Prof. P.S. Letelier and Prof. I. L. Shapiro for helpful discussions on the subject of this paper. I would like to thank CNPq. (Brazil) for partial financial support.

\end{document}